\documentclass{biophys-new}
\usepackage[utf8]{inputenc}
\usepackage{graphicx}
\usepackage[colorlinks,allcolors=cyan!70!black]{hyperref}
\usepackage{float}

\usepackage{lipsum}

\title{Influence of silicon nanocone on cell membrane self-sealing capabilities for targeted drug delivery - computer simulation study}
\runningtitle{Influence of silicon nanocone on cell membrane self-sealing } %% For page header

\author[1]{Przemysław Raczyński}
\runningauthor{Raczyński et. al.}
\author[1]{Krzysztof Górny}

\author[2]{Beata Marciniak}

\author[3,5]{Piotr Bełdowski}

\author[4]{Thorsten Pöschel}

\author[1,*]{Zbigniew Dendzik}

\affil[1]{Faculty of Science and Technology, University of Silesia in Katowice, 75 Pułku Piechoty 1A, 41-500 Chorzow, Poland}
\affil[2]{Faculty of Telecommunications, Computer Science and Electrical Engineering, Bydgoszcz University of Science and Technology, Kaliskiego 7, 85-796 Bydgoszcz, Poland}
\affil[3]{Faculty of Chemical Technology and Engineering, Institute of Mathematics and Physics, Bydgoszcz University of Science and Technology, Kaliskiego 7, 85-796 Bydgoszcz, Poland}
\affil[4]{Institute for Multiscale Simulation, Friedrich-Alexander-Universität Erlangen-Nürnberg, IZNF Cauerstraße 3, 91058, Erlangen, Germany}
\affil[5]{Helmholtz-Zentrum Hereon: Institute for Metallic Biomaterials, Max-Planck-Straße 1, 21502 Geesthacht, Germany}

\corrauthor[*]{zbigniew.dendzik@us.edu.pl}

\papertype{Article}
\begin{document}

\begin{frontmatter}

\begin{abstract}
Efficient and non-invasive techniques of cargo delivery to biological cells are the focus of biomedical research because of their great potential importance for targeted drug therapy. Therefore, much effort is being made to study the characteristics of using nano-based biocompatible materials as systems that can facilitate this task while ensuring appropriate self-sealing of the cell membrane. Here, we study the effects of indentation and withdrawal of nanospear on phospholipid membrane by applying steered molecular dynamics (SMD) technique. Our results show that the withdrawal process directly depends on the initial position of the nanocone. The average force and work are considerably more significant in case of the withdrawal starting from a larger depth. This result is attributed to stronger hydrophobic interactions between the nanocone and lipid tails of the membrane molecules.
Furthermore, when the indenter was started from the lower initial depth, the number of lipids removed from the membrane was several times smaller than the deeper indentation. The choice of the least invasive method for nanostructure-assisted drug delivery is crucial for possible applications in medicine. Therefore, the results presented in this work might be helpful in efficient and safe drug delivery with nanomaterials.
\end{abstract}

\begin{sigstatement}
Efficient and non-invasive methods of nanostructure assisted drug delivery into biological cells are of importance in potential biomedical applications such as targeted drug therapy. Therefore, much effort is being made to study the characteristics of using nano-based biocompatible materials as systems that can facilitate this task while ensuring appropriate self-sealing of the cell membrane. In this work we study the effects of indentation and withdrawal of nanospear on phospholipid membrane by applying steered molecular dynamics (SMD) technique. Accurate characterization of this process is vital in view of introducing novel methods with the use of carbon based nanomaterials.

\end{sigstatement}
\end{frontmatter}

\section*{Introduction}

Nano-structured materials are of great interest in a wide scope of biomedical applications. These applications range from medical imaging techniques and diagnostics to intracellular drug delivery in targeted therapy through direct interfacing nanomaterials with cells. This appears to be an up-and-coming concept due to its direct procedures~\cite{farokhzad2009, riehemann2009, santos2013, santos2012,CAODURO201736}. 
Existing methods applied in such therapies use viruses, electric fields or harsch chemicals and are expensive or toxic or otherwise inefficient, which is why developing improved or novel methods in this field is highly desirable~\cite{Xu2018}. Methods of membrane disruption intracellular delivery, either in vitro or ex vivo, and include electroporation~\cite{xie2013}, squeezing~\cite{ding2017}, fluid shear~\cite{stewart2016}, direct microinjection~\cite{Yoo2013}, photothermal~\cite{Wu2016}, sharp nanostructures~\cite{Yoo2013} and combinations~\cite{ding2017, xie2013}. As an example, configurations of needle-like nanostructures can physically penetrate flexible cell membranes to deliver biomolecules to cells efficiently and in parallel~\cite{Peng2014, Chiappini2015} with minimal impact on viability and metabolism due to their nanoscale sharp tips~\cite{Xu2018}.

Currently, much effort is devoted to studying the interaction between biomembranes and nanostructures \cite{Luna2018,JIANG2017771,Baotong2014}, among others, nanoindentation of the membranes with a number of low-dimensional nanostructures - nanotubes and/or 2D materials~\cite{RaczBBA2018,RaczJPC2019,RaczJPC2020}. Nanocones are considered nanostructures that could mechanically disrupt the cell membranes without causing irreversible membrane damage to cells~\cite{Chong2013}, which might support the delivery of molecules into the cells. In this work, we consider the characteristic of indentation of the biological membrane with a silicone nanocone indenter and its subsequent withdrawal from the membrane to determine how the speed and depth of indentation influence membrane disruption. Although there are several studies on nanocone-like structure indentation onto the membrane \cite{Saavedra2020,Kang2018} none of them looks into velocity dependence and withdrawal processes which may play a critical role in real-life problems when working with nanodevices. Thus, investigating the characteristic of this process is essential to the potential usage of silicon nanocone in the membrane disruption approach of targeted drug delivery methods. On the other hand, highly dispersed CNTs (SWCNTs or MWCNTs) show the ability to adhere to bacterial cells due to van der Waals interactions \cite{BAI20113663,Yang2010}. Furthermore, SWCNTs can not only capture cells but also cause cell death due to the direct physical puncture, resulting in damage to the outer membrane of cells not observed for MWCNTs, probably due to the large diameter of the MWCNTs compared to that of SWCNTs.

Among nano-based materials considered in the biomedical context,  nanostructures based on 2D materials such as graphene, nanotubes, and nanocones, attract much attention ~\cite{BOUKHERROUB202269,CAODURO201736,Baotong2014}. Kim et al. \cite{Yoo2013} showed that mammalian cells could be cultured on vertically aligned arrays of silicon nanowires (SiNW) in such a way that the SiNW naturally penetrates the cells during their growth~\cite{Chong2013,ELNATHAN2014172}. Bioinspired polyethylene terephthalate nanocone arrays with underwater superoleophobicity and anti-bioadhesion properties are also considered promising systems in biomedical applications~\cite{Liu2014}. Precision-guided nanospears are also considered for targeted and high-throughput intracellular gene delivery~\cite{Xu2018}. Innovative solutions based on nanostructures are also considered to enable targeted, rapid, safe, and efficient delivery of biomolecular cargo to target cells and accelerate deployment of new medical interventions to the clinic~\cite{Xu2018}. In the diagnostics of the level of breast cancer-specific protein CA 15-3 in serum a novel and sensitive label-free immunoassay based on gold nanosphere (Au NSs) electrochemically assembled onto thiolated graphene quantum dots (CysA/GQDs) may be used for the detection of CA 15-3 antibodies~\cite{Hasanzadeh2018}.

Biocompatibility and availability of nano-based materials are important factors when constructing bionic devices~\cite{Adabi2017}. A considerable amount of work has been done so far to elaborate on the biocompatibility of SiNW, including some complete toxicological studies; however, most of the data are still insufficient to be conclusive about SiNW biocompatibility~\cite{BOUKHERROUB202269,Marcon2014}. Nevertheless, silicon nanowires (SiNW) are potential candidates for biological applications such as drug delivery systems, in vivo imaging agents and biosensors~\cite{Marcon2014}. Furthermore, as silicon is an abundant element on Earth, silicon-based materials appear to be cost-effective and simultaneously promising materials for nanomedicine, with many putative applications in drug delivery, diagnostics and biosensors~\cite{Marcon2014, santos2012, Das2018}. In addition, this could involve the development of bionic devices for replacing damaged tissues in ears or eyes, which can receive any signal or optical information and convert them into biological signals~\cite{Colpitts2016}.

\section*{Materials and Methods}

MD simulations were performed using the NAMD 2.12~\cite{Schulten2005} application with the all atom-atom CHARMM36~\cite{MacKerell1998,charmm36lipid} force field. VMD 1.9.3~\cite{Humphrey1996} was used as a visualization tool. All the most important parameters of the simulation necessary to recreate it, are prestened in Tab. \ref{MD_param}. The nonbonded parameters for Si (CHARMM atom type Si)  were taken from~\cite{krzem}. All simulations were performed at the physiological temperature T = 310 K and in aqueous environment. TIP3 CHARMM model~\cite{Jorgensen1983} of water, implemented in CHARMM, was used. The water box was explicitly included in the simulations and was modelled as 36612 water molecules. Standard NAMD integrator (Brünger-Brooks-Karplus algorithm)~\cite{Brunger} was used.
The applied model of the phospholipid bilayer is consistent with one used in our earlier simulations~\cite{RaczBBA2018,RaczJPC2019, RaczJPC2020} and consists of 232 1,2-dimyristoyl-sn-glycero-3-phosphocholines (DMPC) and 48 cholesterols. Atomic charges and the model of cholesterol molecules were taken from~\cite{Chipot2006}. The membrane was indented with 69 \AA\ long conical spear with the max diameter equal to 28 \AA, modeled as rigid-body. The system was initially equilibrated in NPT ensemble. The pressure was controlled using Langevin barostat implemented in NAMD. 

The equilibration in NPT ensemble was performed for 2 x ${10}^{6}$ simulation steps. Next, the system was equilibrated in the NVT ensemble (constant number of atoms, constant volume and constant temperature) for the same number of steps. After the equilibration has been completed, the main simulations and collecting data have been performed. During equilibrium stage, the nanospear was kept at the distance of approximately 8 \AA\ from the bilayer surface. During indentation and withdrawal processes, it was moved with the low and high rates (see Tab.~\ref{MD_param} for exact values). This values were chosen to be directly comparable with our previous simulations~\cite{RaczBBA2018, RaczJPC2019, RaczJPC2020} setings\cite{Wallace2008,Pogodin2010,Gangupomu2011}. Periodic boundary conditions were applied with the minimal image convention~\cite{Allen,Frenkel,Rapaport}. The simulation box after NPT equilibration was set to $96 \times 82 \times 176 $ {\AA}. 

Steered Molecular Dynamics (SMD)~\cite{Kale1999, Schulten2005} was used to move nanospear via phospholipid bilayer during indentation and withdrawal processes. Virtual springs were added to all nanospear atoms, which were moved perpendicularly to the bilayer surface with two constant rates. All simulations were repeated to ensure efficient sampling of configuration space. All indentation runs were repeated 15 times and each withdrawal run was repeated five times because of the way of choosing initial runs to withdrawal process (described in the Withdrawal process subsection). Results presented here are averages of these simulation runs. For all main runs, the data necessary to calculate observables presented here were collected every 50 simulation steps. The trajectory was stored every $10^5$ steps.

\begin{table} [hbt!]
  \caption{Simulation parameters}
  \centering
  \begin{tabular}{l l l l}
    \hline
     Description & Value & Unit & Reference  \\
    \hline\\
    $\varepsilon_{Si}$\text & 0.6 &  { kcal } $\text{mol}^{-1}$ & \cite{krzem}\\
    \hline\\
    $\sigma_{Si}$\text & 3.92 & \AA  & \cite{krzem}\\
    \hline\\
    $q_{Si}$ & 0 & e  & \cite{krzem}\\
    \hline\\
    Temperature & 310 & K  & \\
    \hline\\
    Time step & 0.5 & fs  & \\
    \hline\\
    Low nanospear indentation rate & 0.5 & m/s  & \\
    \hline\\
    High nanospear indentation rate & 2 & m/s  & \\
    \hline\\
    Cutoff & 12 & \AA  & \\
    \hline\\
    Langevin thermostat dumping coefficient $\gamma$\text & 1 & ${ps}^{-1}$  &\\
    \hline\\
    Barostat decay time & 100 & ps  & \\
    \hline\\
    Barostat pistion period & 200 & ps  & \\
    \hline\\
    NPT equlibration pressure & 1 & atm  & \\
    \hline\\
    Force constant for SMD spring & 10 & $\text{kcal } \text{mol}^{-1} {
\mathring{\mathrm{A}}^{-2}}$  & \\
    \hline
    \label{MD_param}
  \end{tabular}
\end{table}

\subsection*{Results and discussion}

\subsection*{Indentation process}

\begin{figure}[H]
  \centering
  \includegraphics [scale=0.25] {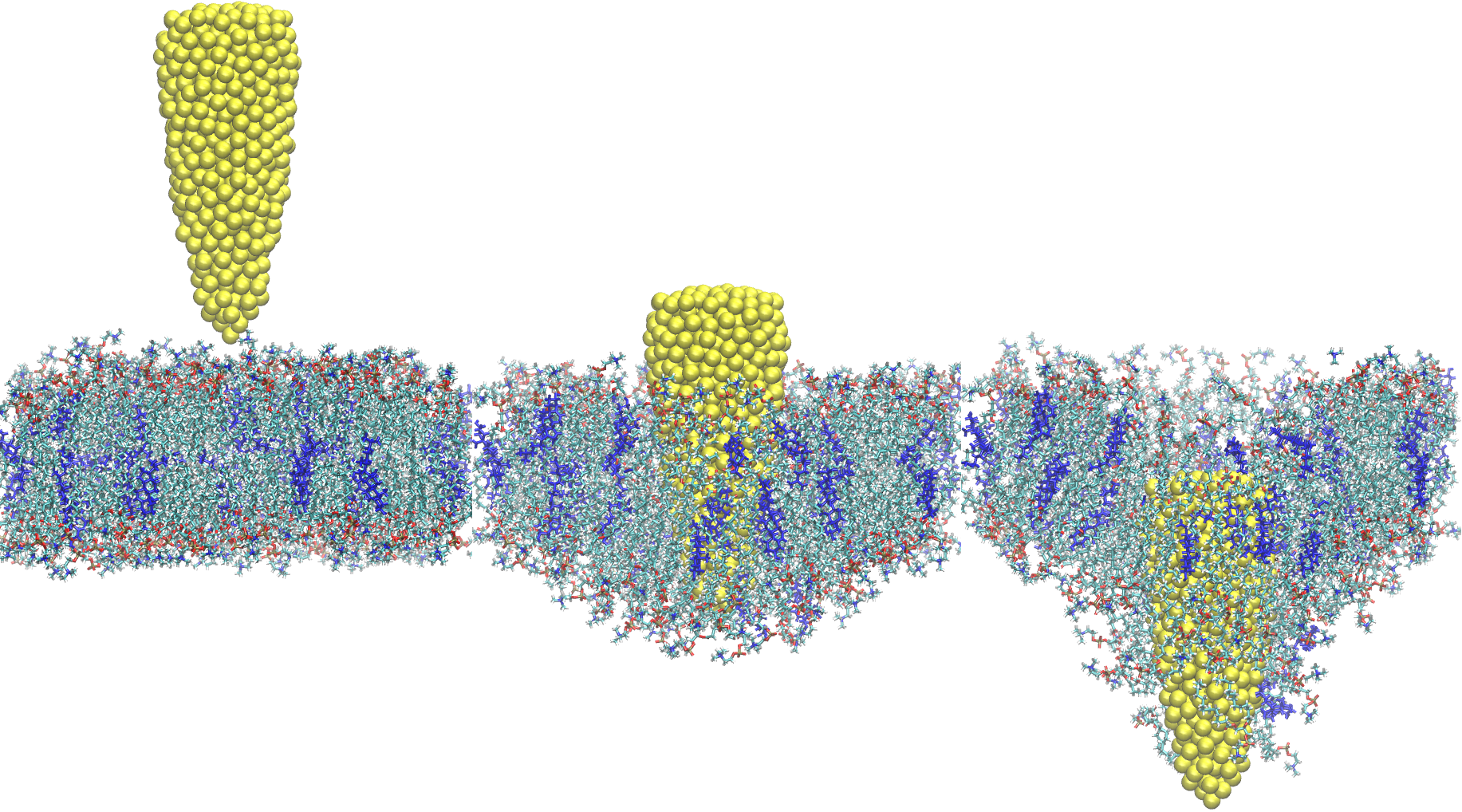}
  \caption{The snapshots of the instantaneous indentation phases: a) initial configuration, b) nanospear travels the distance approximately 63 \AA, c) nanospear travels the distance approximately 105 \AA. The cholesterols in the membrane are colored in blue.}
  \label{in_snapshot}
  \end{figure}

Fig.~\ref{in_snapshot} shows the initial configuration (Fig.~\ref{in_snapshot}a) and instantaneous configurations of the system during indentation process, where the nanospear travels the distance 63 and 105 \AA (Fig.~\ref{in_snapshot}b-c, respectively). The membrane lipids tightly cover the nanospear what suggests that a number of lipids may be removed out of the membrane by the indenter. As a result, a deflection of the membrane referenced to the indentation stage occurs. These observation is discussed later in this work.

\begin{figure}[H]
  \centering
  \includegraphics [scale=0.3] {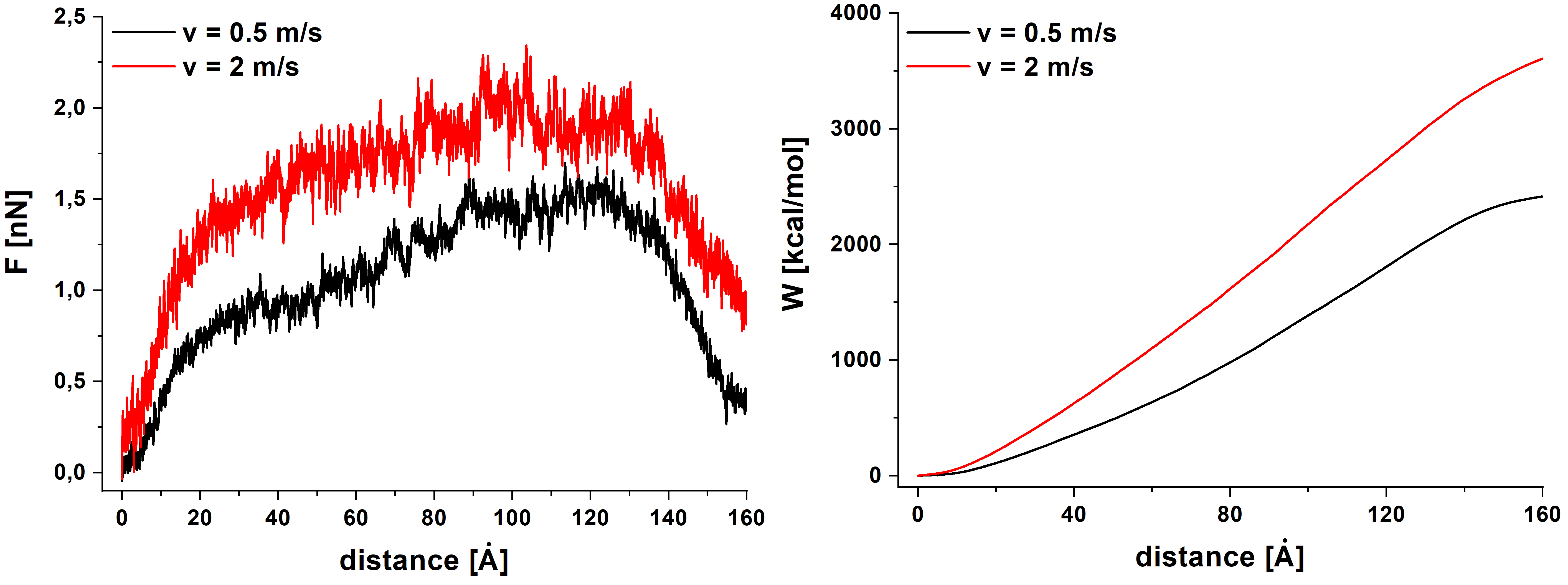}
  \caption{The comparison of average force (left) and work (right) calculated for indentation process with two translation velocities.}
  \label{in_force_work}
  \end{figure}

The force and work required to penetrate the membrane is shown in Fig.~\ref{in_force_work}. The force significantly increases at the initial stage of the indentation process, when the nanospear reaches the membrane surface and starts to barge neighboring DMPC and cholesterol molecules. The value of the force after this stage is similar to the results shown in \cite{RaczBBA2018}, however the force does not diminish when the nanospear barges strong-interacting and rigid glycerol back-bones, as in the case of nanotubes. Contrary, the slope of the force characteristic diminishes, but the force gradually increases as a result of indentation with nanoindenter with diameter increasing in the course of indentation (force characteristic between d = 20-90\AA). The values of force in the mentioned distance regime could be compared to the values of breakthrough force obtained from atomic force microscopy (AFM) experiments for mica supported DOPC - eSM - cholesterol bilayers \cite{James_AFM}. The breakthrough force reported in this study are in the range of 3.0 to 6.5 nN. The shape of the force curves reported in the AFM study is more steep, but this could be associated with the substrate stiffness on which the bilayer is placed. In our study the bilayer is freestanding and can bend, leading to much flatter force characteristic. The discrepancy in maximal value of force required to indent the bilayer can also be attributed to difference in speed of the indenter. In case of computer experiments, it is typical to operate at much higher indentation rates than in case of AFM \cite{James_AFM, guillou_dynamic_2016} due to computational time restrictions. This may also add to the observed difference in the force characteristics \cite{RaczBBA2018}.

In the third region (between 90 and 130 \AA), the force is almost constant. It  decreases significantly when the nanospear indenter leaves the membrane. When the distance traveled by the nanospear is larger than 160 \AA\, the force and work plots become flat. The non-zero values of force results from the water and the lipids removed out of the membrane. These lipids adhere to the nanospear surface and are moving with it through water environment. Because of the just like that shape of the force characteristic, the work required during nanoindentation process is much higher than in the case of nanotubes\cite{RaczBBA2018,RaczJPC2019}. The force and work depends on the speed of indentation - lower rate means lower values of these quantities.
The nanospear strongly interacts with the lipids in the membrane, what manifests itself in large number of lipids removed from the membrane. The nanospear indenter was able to tear out, on average, 42.3 and 30.7 lipids from the membrane (for lower and higher rates, respectively). 15\%\ of molecules permanently removed from the membrane were cholesterols. The number of lipids removed from the membrane was assessed after the nanospear completely left the membrane.
 
\begin{figure}[hbt!]
  \centering
  \includegraphics [scale=0.59] {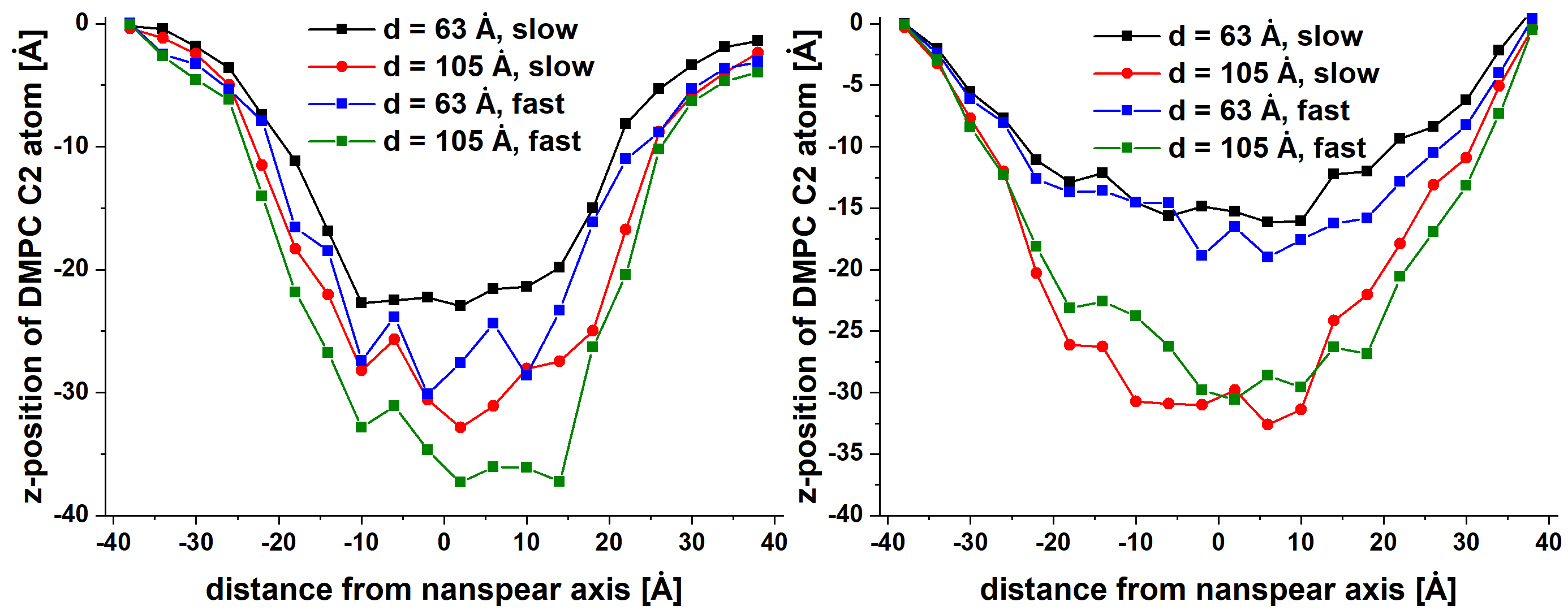}
  \caption{Comparison of average deflection of upper (left) and lower (right) layer. In the legend the letter d means distance travelled by nanospear and the words slow/fast means lower and higher indendation rates. The travelled distances (-63 \AA\ and -105 \AA) correspond to Fig.~\ref{in_snapshot}}. 
  \label{in_deflection}
\end{figure}

\begin{figure}[H]
  \centering
  \includegraphics [scale=0.59] {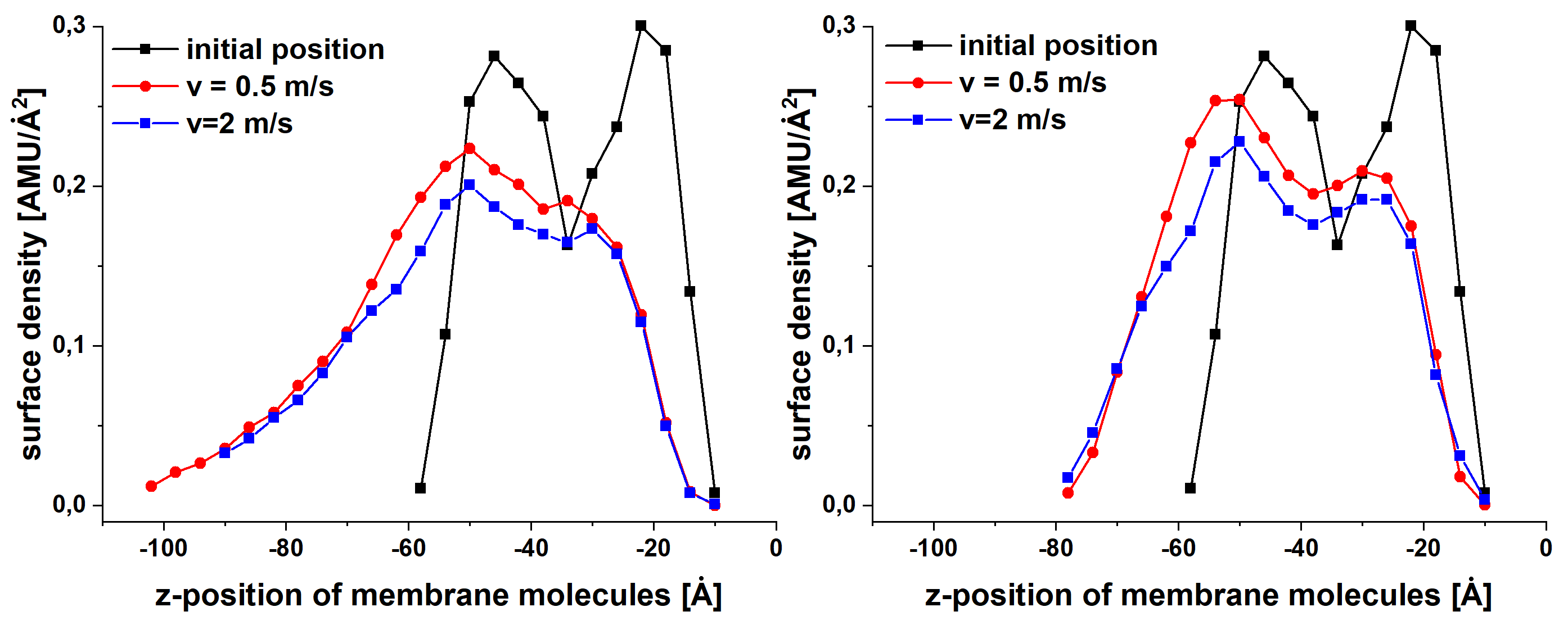}
  \caption{The comparison of the average density profile for the membrane. The travelled distances (-63 \AA\ (left) and -105 \AA (right)) correspond to those shown in Fig.~\ref{in_snapshot}.}.
  \label{in_density}
  \end{figure}

During indentation the membrane experiences deflection, what can be seen in Fig.~\ref{in_snapshot}b-c. The respective plots of average deflection, for upper and lower layer, are shown in Fig.~\ref{in_deflection}. The deflection depends on both, the indenter speed and the displacement. The larger distance traveled corresponds to higher deflection, what is especially pronounced in the case of the bottom layer. Also, higher indentation rate results in higher deflection because it is more difficult for the membrane to accommodate changes caused by fast moving nanospear. Although upper layer experiences larger deflection, for both layers level of deflection results in the changes of membrane structure shown in Fig.~\ref{in_density}. 

Black curves in the average density profile plots correspond to initial positions of the lipids in the membrane. The pressure exerted by the indenter on the upper layer of the membrane affects also its lower counterpart. During the ongoing indentation process, the changes in the structure of the membrane become more pronounced and boundary between the upper and lower layer is more fuzzy (Fig.~\ref{in_density}, right). The clearly visible "tails" in Fig.~\ref{in_density}, right, correspond to these lipids, which adhere to the nanospear and are pulled out of the bilayer (see Fig.~\ref{in_snapshot}c).

\subsection*{Withdrawal process}

\begin{figure}[H]
  \centering
  \includegraphics [scale=0.25] {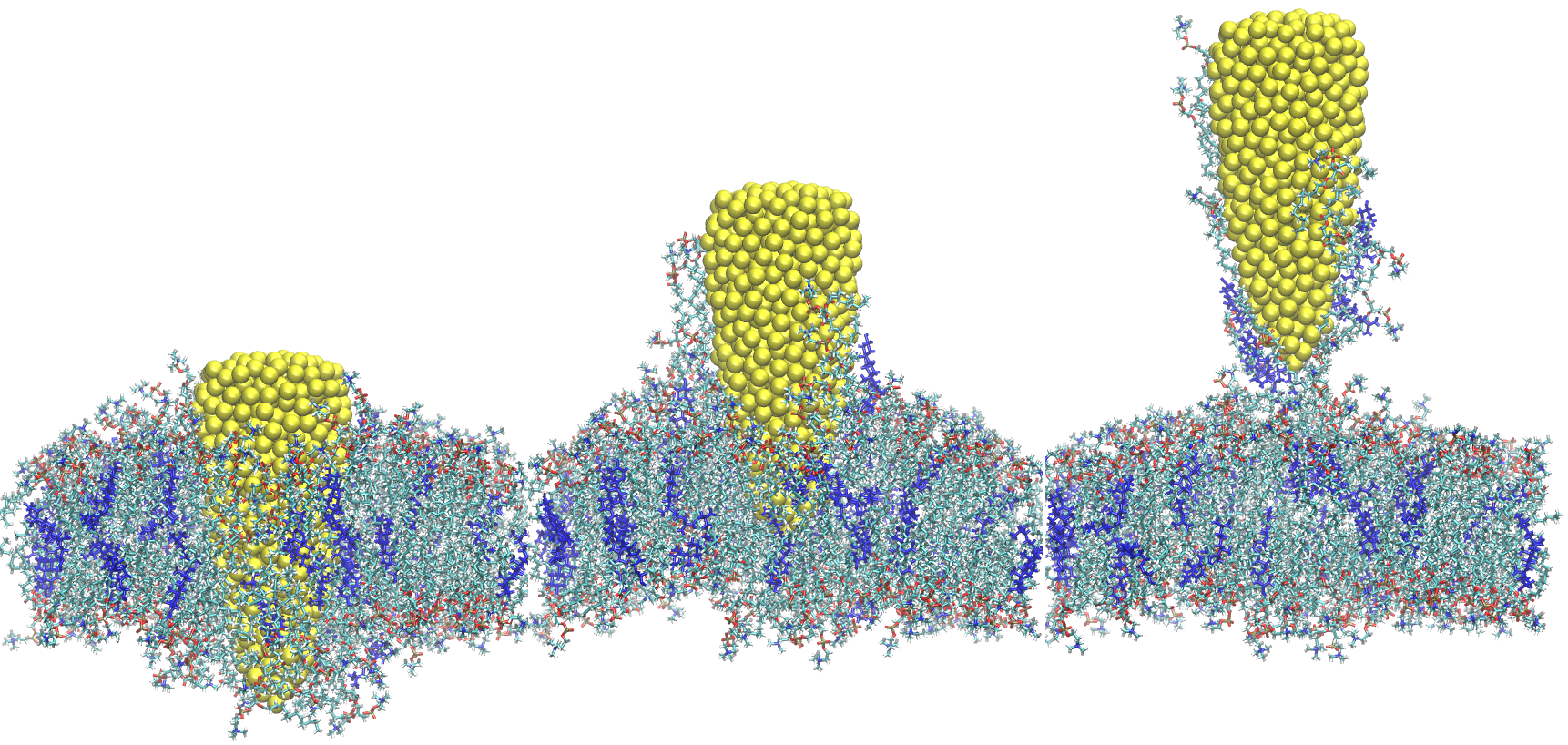}
  \caption{The snapshots of the next withdrawal phases for the system which was started from the larger depth and nanospear was moved with low rate: a) nanospear travels the distance approximately 37 \AA\ (is at the starting position for the smaller depth systems), b) nanospear travels the distance approximately 71 \AA, c) nanospear travels the distance approximately 110 \AA\ (it is 5 \AA\ "above" initial position show in Fig.~\ref{in_snapshot}a).}
  \label{out_screen}
  \end{figure}

After the indentation stage, the configurations of eight systems were chosen as starting configurations to investigate the withdrawal process, four in the case of high indentation rate and four in the case of low indentation rate. In the case of either first set of four or in the case of the second set of four configurations, we selected two trajectories where the number of lipid molecules permanently removed out of the membrane was the highest and two trajectories where this number was close to the average value (given in the Indentation process subsection). The motivation for this method of choosing of the starting configurations for withdrawal process, was to check whether or not the number of lipids removed out of the membrane during indentation phase, influences the characteristic of the withdrawal process. Moreover, to identify possible dependencies between the depth the nanospear reached and the number of removed membrane lipids, two indentation depths were chosen. The smaller one, when nanospear travels the distance, approximately 63 \AA\ during indentation and the larger one, when it travels the distance approximately 105 \AA. The example of starting configurations are shown in Fig.~\ref{in_snapshot}b-c, respectively. The subsequent stages of withdrawal process are shown in Fig.~\ref{out_screen}. During the withdrawal phase, each system was simulated independently five times and the results were averaged over these five trajectories.

Fig.~\ref{out_screen}a shows the structure of the membrane after the displacement of the nanospear equal to 63 \AA\ (the same as in Fig.~\ref{in_snapshot}b for the indentation phase) but now the indenter is pulled up (the withdrawal phase). The comparison with Fig.~\ref{in_snapshot}b implies that the membrane is pulled up by the nanospear. Fig.~\ref{out_screen}c shows that the nanospear indenter is able to remove lipids out of the membrane even with its narrower end. From the Fig.~\ref{out_screen} can be concluded that membrane quite quickly reaches its maximal bulge - its upper layer looks similarly in every stage of the withdrawal process.

\begin{figure}[hbt!]
  \centering
  \includegraphics [scale=0.59] {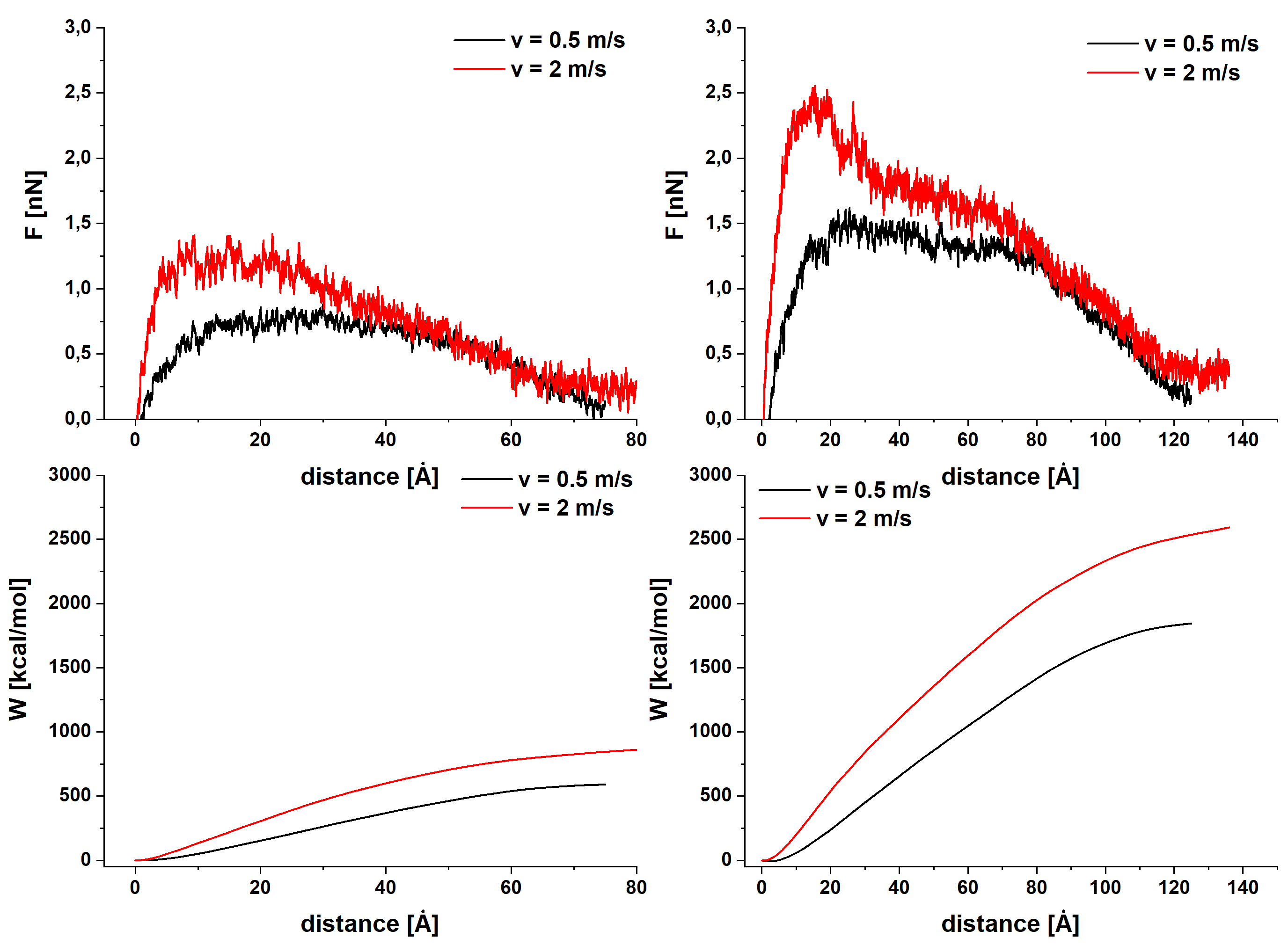}
  \caption{The comparison of the average force (top figures) and work (bottom figures) required during withdrawal process. Left figures show the force and work required when the withdrawal process starts from shallower indentation depth 63 \AA. Right figures show deeper indentation start at 105 \AA. }
  \label{out_force_work}
  
\end{figure}

The first quantities we have calculated were the average force and work versus displacement, during the withdrawal process. Figure ~\ref{out_force_work} shows the comparison of the average force and work during the withdrawal process. However, the obtained results can be slightly affected by the impact of moving the nanospear indenter with the blunt end via water environment. The required force is about two times larger when the withdrawal process starts from the deeper position in the case of both, slower and higher indentation rate. The largest force is required at the beginning of the withdrawal process, when the nanospear has to recoil the membrane. It is particularly visible in the case of the deeper position of the indenter, when it is tightly surrounded by more lipids and is not able to slip away easily from membrane even in the case of larger indentation rate. In each force plot a form of plateau can be observed - it is a stage when the nanospear slides out of the membrane. When the indenter tip starts to leave the membrane the average force quickly is reaching its minimum. The maximal value of force in the case of withdrawal from deeper position (\ref{out_force_work}, top right panel) could be compared with the adhesion force reported in AFM experiment \cite{James_AFM}. The reported experimental values are in the range of 4 to 6 nm and are comparable with our computer experiment data.

Table~\ref{numberremoved1} shows the average number of lipids which were permanently removed from the membrane. These values are significantly lower comparing to the indentation process because the nanostructure during its journey through bilayer becomes more and more narrow. The obtained results suggests that the nanospear should be withdrawn out of the membrane as quickly as possible to minimize amount of molecules removed from the membrane. 

The characteristic of the withdrawal process considerably depends on the initial depth of the nanospear. The difference in depth of 37 \AA\ (see Fig.~\ref{in_snapshot}b,c) implies that the number of molecules removed from the membrane multiplies. On the other hand, the number of molecules removed from the membrane during indentation process does not correlate with the number of molecules removed during the withdrawal phase. For example, in one of the runs of indentation with low rate, 49 lipids were removed from the membrane, what is second worst result (average is equal to 42.3 molecules). This run was chosen as an initial for withdrawal process, however the average number of removed lipids is equal to 8.4, what is lower than the average value (10.8 lipids).

\begin{table} [H]
  \caption{Number of lipids removed during withdrawal process}
  \label{numberremoved1}
  \centering
  \begin{tabular}{l l l}
    \hline
     & Smaller depth & Larger depth  \\
    \hline\\
    low rate & 1.5 (1.3)  & 10.8 (4.1) \\
    \hline\\
    high rate & 0.2 (0.4)  & 5.7 (3.4) \\
    \hline
  \end{tabular}
\end{table}

\begin{figure}[H]
  \centering
  \includegraphics [scale=0.59] {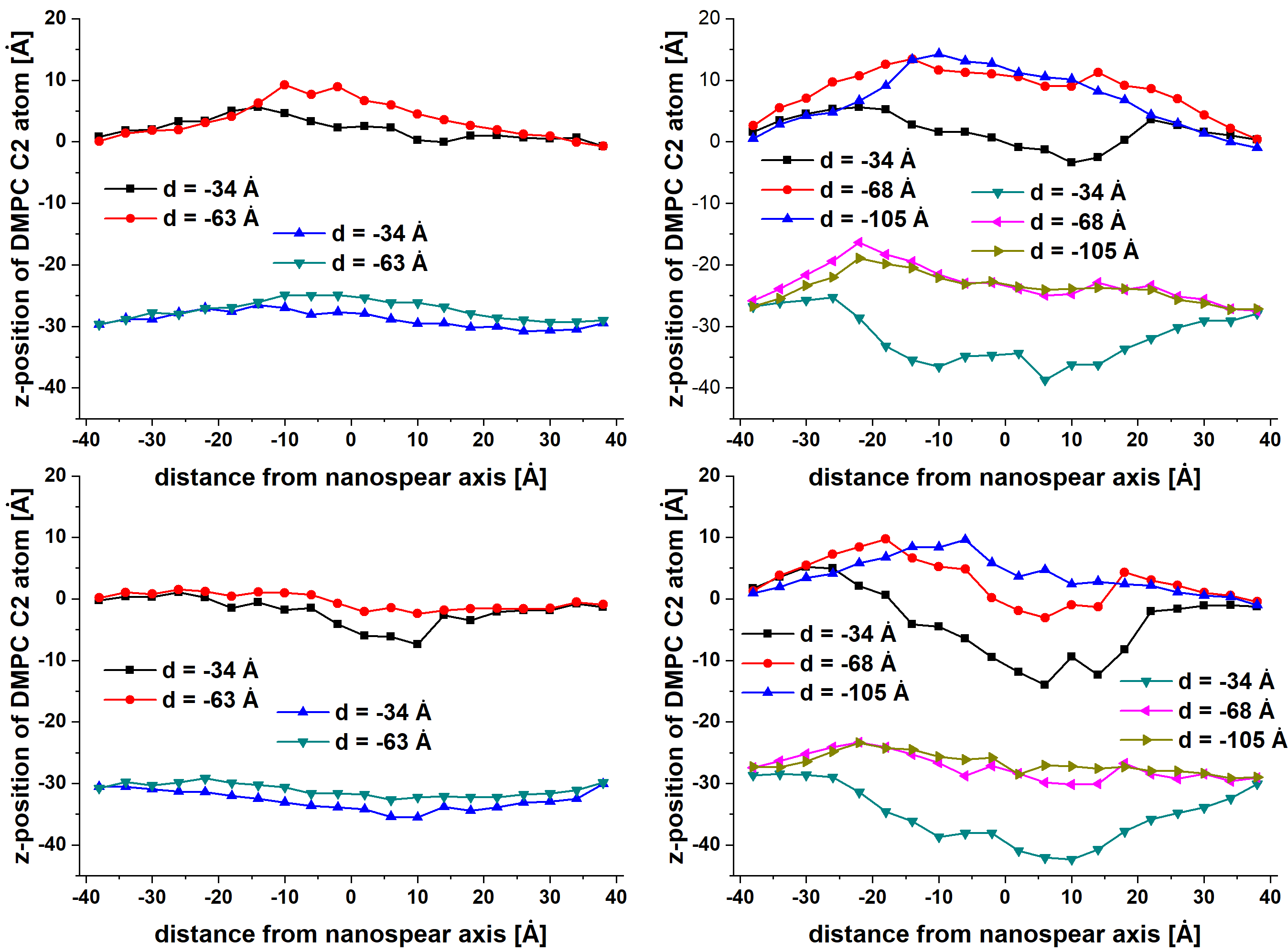}
  \caption{DMPC backbone C2 atom average positions for lower (top figures) and higher (bottom figures) withdrawal rate. Left figures show the C2 atom positions when the withdrawal process starts from shallower indentation depth 63 \AA. Right figures show deeper indentation start at 105 \AA.}
  \label{out_grzybek}
  \end{figure}

During withdrawal process, the membrane is pulled up by nanospear (see Fig.~\ref{out_grzybek}). This effect is more visible for slow withdrawal rate or larger initial depth of the nanostructure. The plots of the average position of DMPC backbone C2 atom confirm the previous conclusions, that the highest possible rate of withdrawal is most desirable. Then, the membrane lipids are no longer able to accommodate changes caused by nanostructure and the membrane is pulled up to smaller extent (Fig.~\ref{out_grzybek}, bottom). Some similarities in bulge of the membrane occur in early/middle stage of the withdrawal (see Fig.~\ref{out_grzybek}, right) for both, low and high rates. At these stages, the nanospear effectively pulls the membrane up. However, at the final stage of withdrawal, the bilayer reaches maximal bulge which is larger for low indentation rate. The level of bulge is significantly smaller comparing to the level of deflection during indentation (compare Fig.~\ref{in_deflection}). 

\begin{figure}[H]
  \centering
  \includegraphics [scale=0.59] {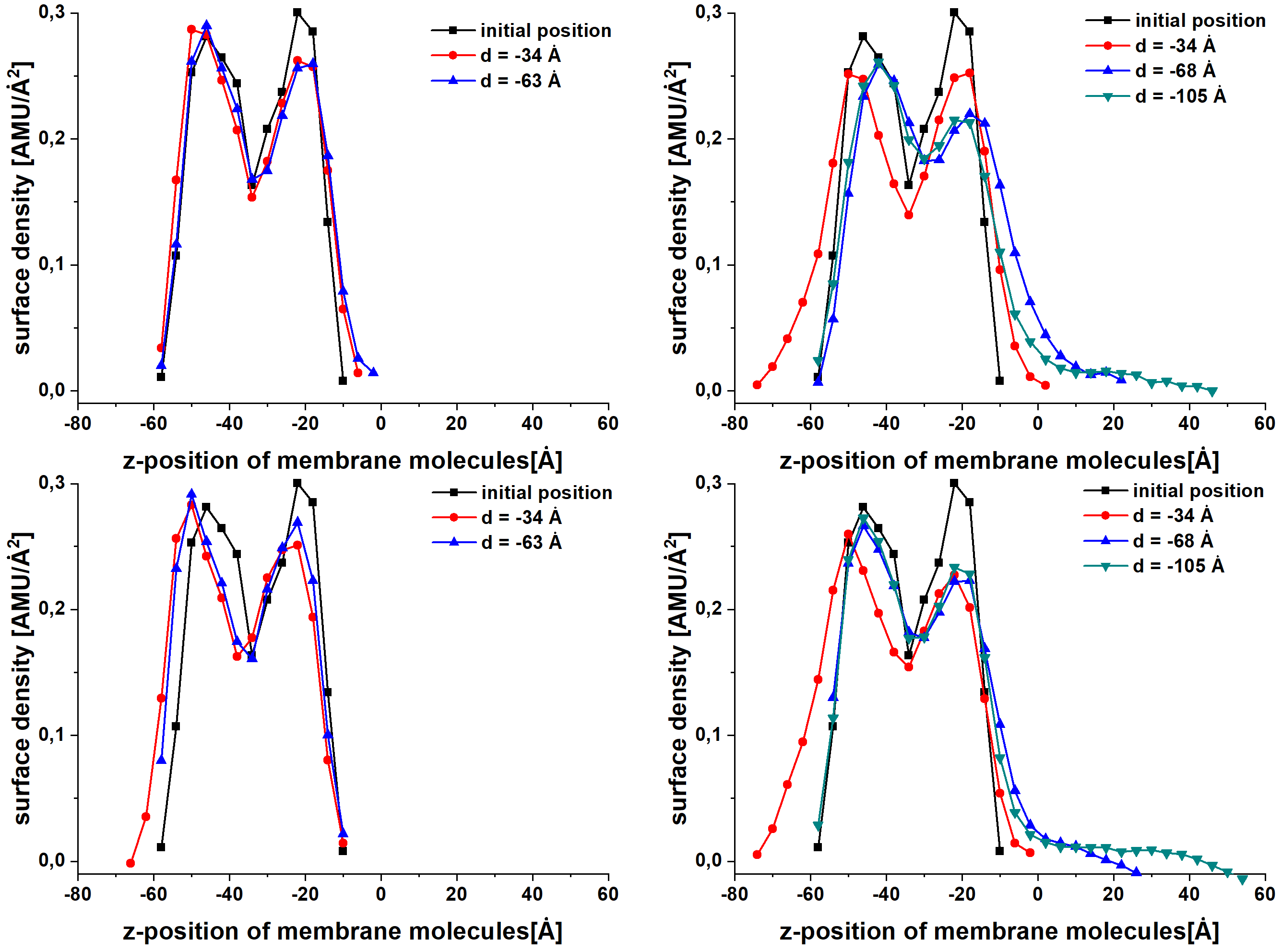}
  \caption{Average density profile for lower (top figures) and higher (bottom figures) withdrawal rate. Left figures show density profile when the withdrawal process starts from shallower indentation depth 63 \AA. Right figures show deeper indentation start at 105 \AA. 'initial position' in the legend means configuration also before indentation process.}
  \label{out_density}
  \end{figure}

Figure~\ref{out_density} shows the average density profile of the membrane undergoing the withdrawal. Note, that the initial configuration here is the configuration of the system after equilibration (i.e. before indentation and withdrawal processes). In the case when the nanospear is at a smaller depth (Fig.~\ref{out_density}, left), not significant changes occurs, when comparing to the initial configuration. More pronounced differences occur when the withdrawal process starts from the larger depth, see Fig.~\ref{out_density}, right. After the nanospear penetrates the initial 37 \AA\ (red curve, right figures), the mass distribution of the membrane most significantly differs from the initial distribution. After the nanospear travelled the distance 71 \AA\ and, subsequently, 39 \AA, the differences become less pronounced. It confirms the previous conclusion that the last stage of withdrawal process exerts only slight effect on the phospholipid bilayer. Blue and olive "tails" in Fig.~\ref{out_grzybek} reflects these molecules which were removed from the membrane by the nanospear indenter. 
  
Instantaneous configuration shown in Fig.~\ref{out_screen}c, small (comparing to indentation) number of molecules removed from the bilayer (Table~\ref{numberremoved1}), slight bulges (Fig.~\ref{out_grzybek}) and similarities in density profile to the initial mass distribution indicates high ability of the membrane to self-sealing after indentation and withdrawal processes. High affinity to self-sealing was also reported, in the case of the indentation of the phospholipid bilayer with carbon nanotubes\cite{RaczBBA2018}, silicon-carbide nanotubes\cite{RaczJPC2019} or graphene\cite{RaczJPC2020} and shows the efficiency of the membrane as a barrier protecting interior of membrane from the outside environment.

\section*{Conclusion}

The main focus of the presented research is to investigate the nanoindentation withdrawal process with emphasis on the initial depth of the nanospear indenter. The results discussed here show that the withdrawal process directly depends on the initial position of the nanospear. The average force and work, are considerably larger in case of the withdrawal starting from larger depth. The differences are also clearly visible in the density profile and in the characteristic of the average positions of lipid molecules. However, the most significant distinction occurs in the number of molecules removed from the membrane. When the indenter has been started from the lower initial depth, the number of lipids removed out of the membrane was several times smaller comparing it to the deeper case. 

The results also show high self-sealing abilities of the membrane after interference of the nanospear on its structure. It increases as the nanoidenter taper gradually leaves the membrane giving the bilayer time to accommodate the changes caused by nanospear during indentation and withdrawal processes. It is important that self-sealing process appears almost immediately and very slightly depend on the interference rates studied.

The withdrawal process, as well as indentation, depends on the rate of the indenter. Taking into account damages caused by moving nanospear during indentation process, two competitive processes occurs. When the membrane is indented with low rate, it is able to tear out larger number of molecules but the deflection of the membrane is smaller. When the indenter was moved with the high rate, the number of lipids removed from the bilayer is smaller but its deflection is larger. In the case of the withdrawal process, the rate should be high because of the smaller number of lipids removed out of the membrane, which is particularly important when it comes to the potential medical applications. Also the bulge of the bilayer is smaller in case of higher rate. 

The use of a silicon nanospear structure, that is more susceptible to bio-toxitcity reducing modifications than the carbon nanotubes, has more importance from the point of view of potential medical applications. It should be also noted, that although the diameter of nanocone is significantly higher (almost two times) than in case of largest nanotube previously studied, the deflection of the membrane during removal process is comparable. Also the number of molecules extracted from the membrane is similar, although it must be taken into account that when removing indenter from a greater depth, the number of phospholipids removed increases significantly in case of silicon nanocone. The application of silicon nanospear is also desirable in comparison to the multi walled carbon nanotubes, which are currently available as indenter needles for soft matter, due to their shape and smaller effective diameter. Further studies could focus on experimental confirming of our results and testing the effects of arrays of nanostructures on biomembranes or molecular aggregates like liposomes or lipoproteins.

\section*{Author Contributions}

% Author1 designed the research. Author2 carried out all simulations, analyzed the data. Author1 and Author2 wrote the article. 
P.R. designed the research. P.R and K.G. designed the simulation protocol. P.R and Z.D performed the simulations. P.R., K.G., B.M., P.B. and Z.D. analyzed the data. P.R., Z.D., P.B and T.P. wrote the article.

% \section*{Acknowledgments}

\section*{Declaration of Interest}

The authors declare no competing interests.

% We thank G. Harrison, B. Harper, and J. Doe for their help.

% Uncomment if using bibtex (default)
\bibliography{refs}

% Uncomment if using biblatex
% \printbibliography

% \section*{Supplementary Material}

% An online supplement to this article can be found by visiting BJ Online at \url{http://www.biophysj.org}.

\end{document}